\documentclass[aps,prl,showpacs,twocolumn,superscriptaddress]{revtex4} 
\usepackage{graphicx}% Include figure files
\begin{document}

\title{
First-principles study on superconductivity of P- and Cl-doped H$_3$S
}

\author{Akitaka Nakanishi}%
\email{nakanishi@hpr.stec.es.osaka-u.ac.jp}
\affiliation{%
  Center for Science and Technology under Extreme Conditions,
  Graduate School of Engineering Science, Osaka University,
  1-3 Machikaneyama, Toyonaka, Osaka 560-8531, Japan
}%

\author{Takahiro Ishikawa}%
\affiliation{%
  Center for Science and Technology under Extreme Conditions,
  Graduate School of Engineering Science, Osaka University,
  1-3 Machikaneyama, Toyonaka, Osaka 560-8531, Japan
}%

\author{Katsuya Shimizu}%
\affiliation{%
  Center for Science and Technology under Extreme Conditions,
  Graduate School of Engineering Science, Osaka University,
  1-3 Machikaneyama, Toyonaka, Osaka 560-8531, Japan
}%

\date{\today}% It is always \today, today,
%\date{September 10, 2005}

\begin{abstract}
%The maximum length is 600 characters including spaces.
The recent reports on 203\,K superconductivity in compressed hydrogen sulfide, H$_3$S,
has attracted great interest in sulfur-hydrogen system under high pressure.
Here, we investigated the superconductivity of P-doped and Cl-doped H$_3$S
 using the first-principles calculations based on the supercell method,
 which gives more reliable results on the superconductivity in doped systems
 than the calculations
 based on the virtual crystal approximation reported earlier. 
The superconducting critical temperature
 is increased from 189 to 212\,K at 200\,GPa in a cubic $Im\bar{3}m$ phase
 by the 6.25\% P doping, whereas it is decreased to 161\,K by the 6.25\% Cl doping.
Although the Cl doping weakens the superconductivity,
 it causes the $Im\bar{3}m$ phase to be stabilized in a lower pressure region than that in the non-doped H$_3$S.

\end{abstract}

% https://ufn.ru/en/pacs/
%
% 63.00.00 Lattice dynamics
% 63.20.dk First-principles theory
%
% 74.00.00 Superconductivity
% 74.62.Dh Effects of crystal defects, doping and substituion
% 74.62.Fj Effects of pressure
% 74.70.-b Superconductiong materials other than cuprates

%\keywords{}

\maketitle

%\section{}
%\subsection{}
%\subsubsection{}

\section{Introduction}
Hydrogen sulfide (H$_2$S) shows a high-temperature superconductivity under high pressure,
 and the superconducting critical temperature ($T_\mathrm{c}$) reaches
 the maximum of 203\,K at pressure of 155\,GPa,
 in which H$_2$S transforms into a stoichiometric compound 
with a chemical formula of H$_3$S
 via the intermediate compounds  
 \cite{Drozdov2015a, Li2014, Li2016, Duan2014, Errea2016, Einaga2016, Ishikawa2016, Akashi2016}.
The mechanism of the high-$T_\mathrm{c}$ superconductivity is considered to have origins of electron-phonon interaction
 because the isotope effect on the superconducting transition was experimentally observed,\cite{Drozdov2015a} which 
supports Ashcroft's prediction 
that metallic hydrides become high-$T_\mathrm{c}$ conventional superconductors.\cite{Ashcroft2004}
The conventional superconductors have an advantage that the $T_\mathrm{c}$
 can be predicted more quantitatively by the first-principles calculation
 than that in copper oxide superconductors and iron-based ones. 
Therefore, 
 a cooperation between experimental measurements and first-principles calculations
 is effective for the exploration of novel superconducting hydrogen compounds.

Superconductivity has been observed experimentally in only a few hydrogen-containing compounds,
 namely silane (SiH$_4$) with a $T_\mathrm{c}$ of 17\,K at 96\,GPa \cite{Eremets2008}
 and phosphine (PH$_3$) with a $T_\mathrm{c}$ of 100\,K at 207\,GPa,\cite{Drozdov2015b}
 in addition to H$_2$S.
However, superconductivity is predicted in many other compounds
 using first-principles calculations (Table \ref{tab:tc_hydrides}).
More information is summarized in
 Refs. \cite{Peng2017, Duan2017, Wang2017, Zurek2017, Zhang2017}.

\begin{table}[htbp]
\caption{
Superconducting critical temperatures predicted
 from first-principles calculation in hydrogen compounds.
 }
\begin{tabular}{cccc}
\hline
             & P (GPa) & $T_\mathrm{c}$ (K) & Reference  \\ \hline
  YH$_{10}$  & 400     & 303                & \cite{Peng2017}      \\
  LaH$_{10}$ & 210     & 286                & \cite{Liu2017}      \\
  MgH$_6$    & 400     & 271                & \cite{Feng2015}     \\
  YH$_6$     & 120     & 264                & \cite{Li2015}       \\
  CaH$_6$    & 150     & 235                & \cite{Wang2012}     \\
  SiH$_4$    & 202     & 166                & \cite{Feng2006}     \\
  AsH$_8$    & 450     & 151                & \cite{Fu2016}       \\
  AlH$_5$    & 250     & 146                & \cite{Hou2015}      \\
  BiH$_5$    & 300     & 119                & \cite{Ma2015b}      \\
  SbH$_4$    & 150     & 118                & \cite{Ma2015a}      \\
  BiH$_6$    & 300     & 113                & \cite{Ma2015b}      \\
  SiH$_8$    & 250     & 107                & \cite{Li2010}       \\ 
  PbH$_8$    & 230     & 107                & \cite{Cheng2015}    \\
  ArH$_4$    & 1500    &  72                & \cite{Ishikawa2017} \\
\hline
\end{tabular}
\label{tab:tc_hydrides}
\end{table}
% reference
% table1/scf.findsym.txt        <- ~/H24S7P/3vcrelax/dope/200GPa/scf.in 
% table1/P12.5.opt.findsym.txt  <- ~/H24S7P/3vcrelax/dope/200GPa/opted.in 
% table1/Cl12.5.opt.findsym.txt <- ~/H24S7Cl/3vcrelax/dope/200GPa/opted.in 

Intentional introduction of impurities into substances
 has been used as another approach
 for the exploration of high-$T_\mathrm{c}$ superconducting materials. 
Ge \textit{et al.} have investigated
 the effect of impurity doping on the superconductivity of H$_3$S
 using first-principles calculations
 based on the virtual crystal approximation (VCA)
 and predicted that the $T_\mathrm{c}$ shows a further increase to
 280\,K at 250\,GPa by 7.5\% substitution of phosphorus (P) for S, 
\textit{i.e.}, hole doping.\cite{Ge2016}
The hole doping causes an increase in the density of states (DOS) at the Fermi level,
 which results in the enhancement of the $T_\mathrm{c}$.
At almost the same time,
 Fan  \textit{et al.} also reported
 the doping effect on the superconductivity of H$_3$S
 using VCA calculations.\cite{Fan2016}
Their calculations show
 an increase of $T_\mathrm{c}$ from 185 to 197\,K at 200\,GPa
 by 15\% P doping,
 which is due to the increase of electron-ion matrix elements. 

In this paper,
 we report the results of the doping effect
 on the superconductivity of H$_3$S,
 obtained by first-principles calculations
 based on the supercell method,
 which gives more reliable results on the superconductivity in doped systems
 than the VCA method. 
Comparing the VCA results reported earlier,\cite{Ge2016,Fan2016}
 we notice that the amount of the increase of $T_\mathrm{c}$ 
 is significantly different between them:
 194 to 250\,K in Ge's calculations
 and 185 to 187\,K in Fan's calculations in 10\% P-doped H$_3$S at 200\,GPa. 
Therefore,
 we verified the VCA results applying the supercell method to
 the $T_\mathrm{c}$ calculations for P-doped
 and chlorine (Cl)-doped H$_3$S.

\begin{table*}[t]
\caption{
Crystal coordinates before and after the structure optimizations
 for 12.5\%-doped $Im\bar{3}m$ H$_3$S at 200\,GPa. 
 }
\begin{tabular}{rrccccccccc}
\hline
   & &\multicolumn{3}{c}{Before}  & \multicolumn{3}{c}{After (P doping)}  & \multicolumn{3}{c}{After (Cl doping)} \\ \hline
      H1 & 12d & 0.25    & 0       & 0.5     & \multicolumn{3}{c}{no change} & \multicolumn{3}{c}{no change}  \\
      H2 & 24h & 0       & 0.25    & 0.25    & 0       & 0.24939 & 0.24939   & 0       & 0.24853 & 0.24853    \\
      H3 & 12e &-0.25    & 0       & 0       &-0.24494 & 0       & 0         &-0.26885 & 0       & 0          \\
      S1 &  8c & 0.25    & 0.25    & 0.25    & \multicolumn{3}{c}{no change} & \multicolumn{3}{c}{no change}  \\
      S2 &  6b & 0       & 0.5     & 0.5     & \multicolumn{3}{c}{no change} & \multicolumn{3}{c}{no change}  \\
  P(Cl)1 &  2a & 0       & 0       & 0       & \multicolumn{3}{c}{no change} & \multicolumn{3}{c}{no change}  \\
\hline
\end{tabular}
\label{tab:structure}
\end{table*}
% reference
% table1/scf.findsym.txt        <- ~/H24S7P/3vcrelax/dope/200GPa/scf.in 
% table1/P12.5.opt.findsym.txt  <- ~/H24S7P/3vcrelax/dope/200GPa/opted.in 
% table1/Cl12.5.opt.findsym.txt <- ~/H24S7Cl/3vcrelax/dope/200GPa/opted.in 

\section{Methods}

\begin{figure}[htbp]
\begin{center}
  \begin{tabular}{c}
  \includegraphics[width=8.5cm]{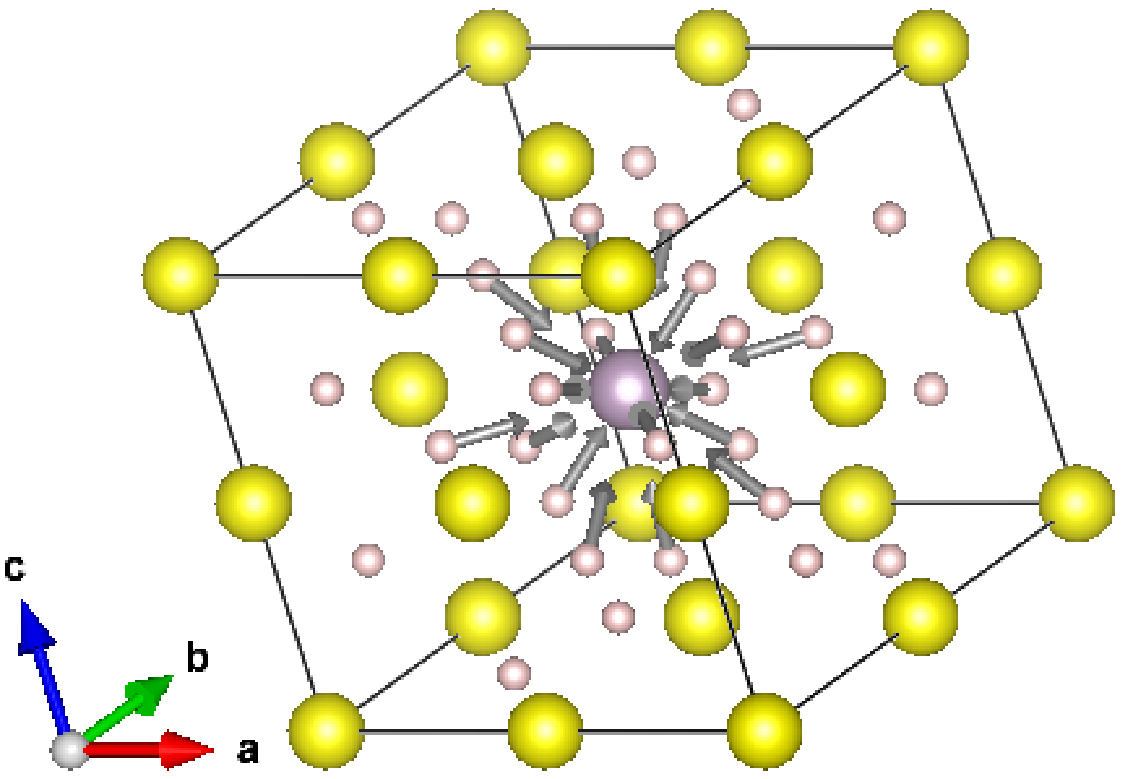} \\
  \includegraphics[width=8.5cm]{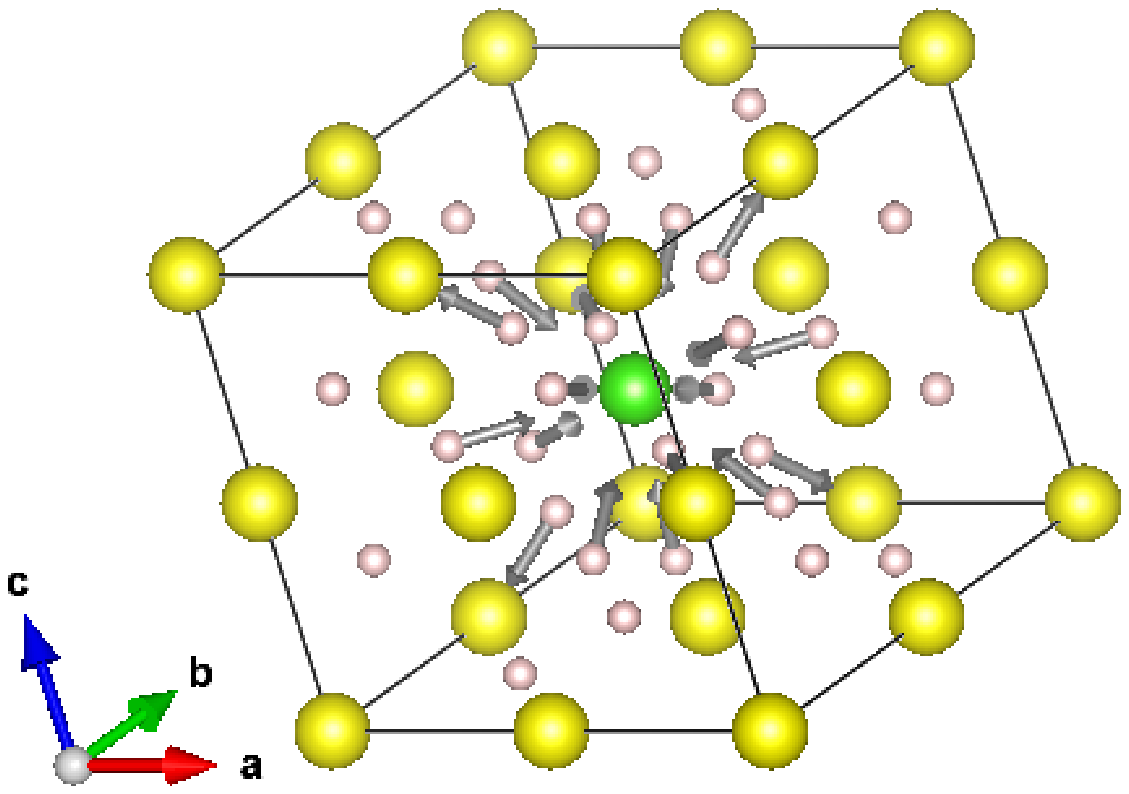} 
  \end{tabular}
  \caption{
Optimized structures of H$_3$S$_{0.875}$P$_{0.125}$ (upper) and 
 H$_3$S$_{0.875}$Cl$_{0.125}$ (lower). 
A large gray (green) sphere at center, large yellow spheres, and small pink spheres
 represent P (Cl), S, and H atoms, respectively. 
Arrows indicate the displacements of the H atoms
 from the equilibrium positions in non-doped H$_3$S. 
The figure produced using the VESTA software package.\cite{Momma2011}
}
  \label{fig:structure}
\end{center}
\end{figure}

We investigated the pressure dependence of superconductivity of H$_3$S
 at 6.25\% and 12.5\% doping of P (or Cl).
First we prepared
 (A) a supercell consisting of $2\times2\times2$ primitive cells,
 including 8 formula units (f.u.) of H$_3$S,
 and (B) that consisting of $2\times2\times2$ conventional cells,
 including 16\,f.u..
Next we created doped H$_3$S
 by substituting 12.5\% of S atoms
 in the supercell A with P or Cl,
\textit{i.e.} H$_3$S$_{0.875}$P$_{0.125}$ and H$_3$S$_{0.875}$Cl$_{0.125}$, 
 and 6.25\% in the supercell B, 
\textit{i.e.} H$_3$S$_{0.9375}$P$_{0.0625}$ and H$_3$S$_{0.9375}$Cl$_{0.0625}$. 
Then we performed structural optimization for the supercells
 in the pressure region of 100 to 250\,GPa
 using the Parrinello-Rahman method without constraint of symmetry.\cite{Parrinello1980}
Comparing the optimized structures with that of the non-doped H$_3$S, 
we found that 
only H atoms are slightly moved from the starting positions in crystal coordinates 
 by the optimization. 
The results at 200\,GPa are listed in Table \ref{tab:structure}
 and are illustrated in Fig. \ref{fig:structure}.

We performed first-principles calculations
 using the Quantum ESPRESSO code,\cite{Giannozzi2017}
 in which the plane wave basis and pseudopotential methods are employed. 
We adopted the Vanderbilt type ultrasoft pseudopotential \cite{Vanderbilt1990}
 and a generalized gradient approximation
 of the Perdew-Burke-Ernzerhof type
 for the exchange correlation functional.\cite{Perdew1996}
These pseudopotentials 
 are available as H.pbe-van\_bm.UPF, S.pbe-van\_bm.UPF, P.pbe-van\_ak.UPF, and Cl.pbe-n-van.UPF on Quantum ESPRESSO pseudopotential library.\cite{QEPPlibrary}
The integration of reciprocal lattice space
 was performed using the Monkhorst-Pack grid
 with a broadening parameter of 0.01\,Ry.\cite{Monkhorst1976}
The numbers of the grids are summarized in Table \ref{tab:kpts}.
The energy cutoff of the wave function was set at 80\,Ry. 

\begin{table}
\caption{
 $k$-point and $q$-point grids used for the calculations.
 }
\begin{tabular}{cccc}
\hline
   Doping amount & 0\% & 6.25\% & 12.5\% \\ \hline
   Electron        & $16\times16\times16$ & $ 8\times 8\times 8$ & $ 8\times 8\times 8$ \\
   Phonon          & $ 4\times 4\times 4$ & $ 2\times 2\times 2$ & $ 4\times 4\times 4$ \\
   Electron-phonon & $32\times32\times32$ & $16\times16\times16$ & $32\times32\times32$ \\
\hline
\end{tabular}
\label{tab:kpts}
\end{table}

For superconductivity, we calculated
 the dynamical matrix,
 phonon frequency,
 electronic phonon matrix,
 and Eliashberg function $\alpha^2 F(\omega)$
 using density functional perturbation theory.\cite{Baroni2001}
The $T_\mathrm{c}$ was calculated
 using the Allen-Dynes-modified McMillan formula,\cite{McMillan1968, Allen1975}
\begin{eqnarray}
  T_\mathrm{c}&=&\frac{f_1 f_2 \omega_{\log}}{1.2}
  \exp \left[ \frac{-1.04(1+\lambda )}
  {\lambda-\mu^{\ast}(1+0.62\lambda )} \right].
\end{eqnarray}
Here,
 $\lambda$ is the electron-phonon coupling constant,
 $\omega_{\log}$ is the logarithmic averaged phonon-frequency,
 $\mu^{\ast}$ is the screened Coulomb interaction constant, and
 $f_1$ and $f_2$ are correction factors for the system showing large $\lambda$.
The value of $\mu^{\ast}$ is assumed to be 0.13,
 which holds for metallic hydrides.\cite{Sanna2018}
The other parameters,
 $f_1$, $f_2$, $\lambda$, and $\omega_{\log}$
 are defined as follows: 
\begin{eqnarray}
  f_1
  &=&
  \left\{1+\left[\frac{\lambda}{2.46(1+3.8\mu^{\ast})} \right]^{3/2}\right\}^{1/3}, \\
  f_2
  &=&
  1+\frac{(\omega_2/\omega_{\log}-1)\lambda^2}
  {\lambda^2+[1.82(1+6.3\mu^{\ast})(\omega_2/\omega_{\log})]^2}, \\
  \lambda
  &=& 2 \int_0^\infty
  d\omega\frac{\alpha^2 F(\omega)}{\omega},\\
  \omega_{\log} 
  &=&\exp \left[\frac{2}{\lambda}\int_0^\infty
  d\omega\frac{\alpha^2F(\omega)}{\omega}\log\omega\right], 
\end{eqnarray}
where $\omega_2$ is defined as 
\begin{eqnarray}
  \omega_2
  &=& \left[\frac{2}{\lambda}\int_0^\infty
  d\omega\alpha^2F(\omega)\omega\right]^{1/2}.
\end{eqnarray}
The calculated $T_\mathrm{c}$ values were improved
 using the superconducting density functional theory \cite{Flores-Livas2016}
 and the inclusion of anharmonic effect on phonon.\cite{Errea2015} 
However, 
 we discuss the superconductivity without these improvements
 to compare our results with the VCA results reported earlier.\cite{Ge2016,Fan2016}

% reference
% table1/P12.5.opt.vector2.png
% table1/Cl12.5.opt.vector2.png

\section{Results and discussion}

% \begin{widetext}
\begin{figure}[htbp]
\begin{center}
  \begin{tabular}{c}
  \includegraphics[width=8.5cm]{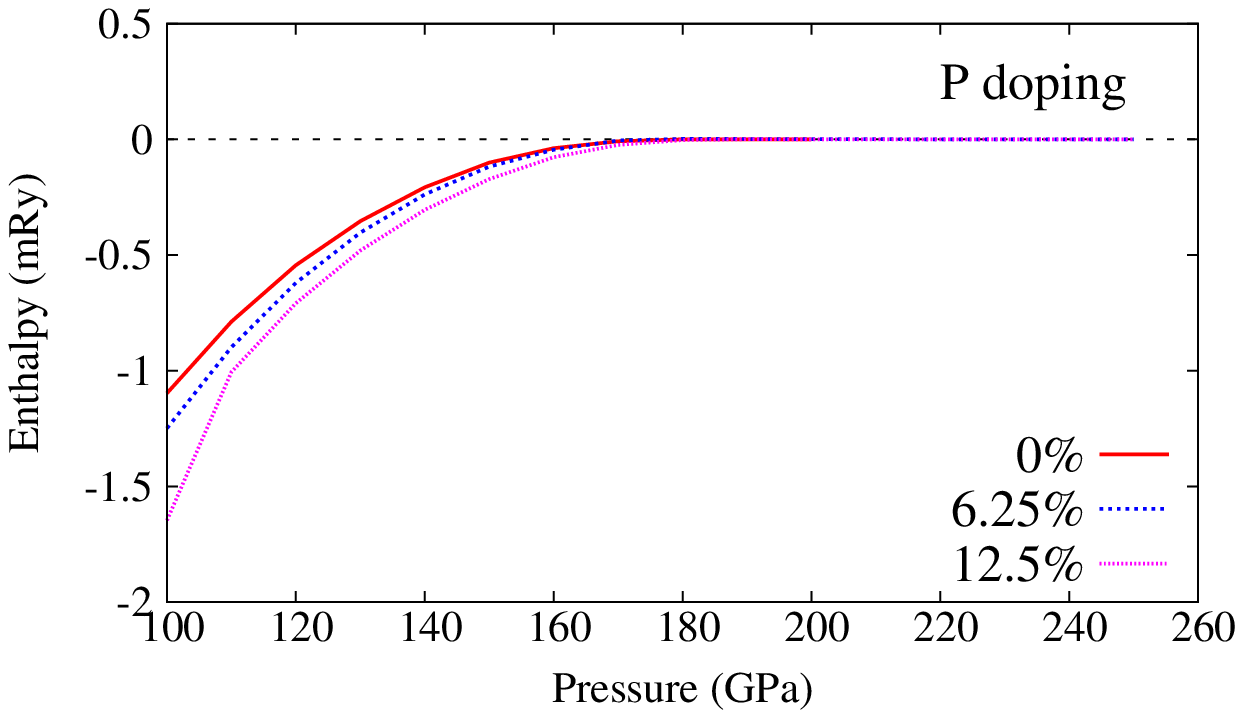} \\
  \includegraphics[width=8.5cm]{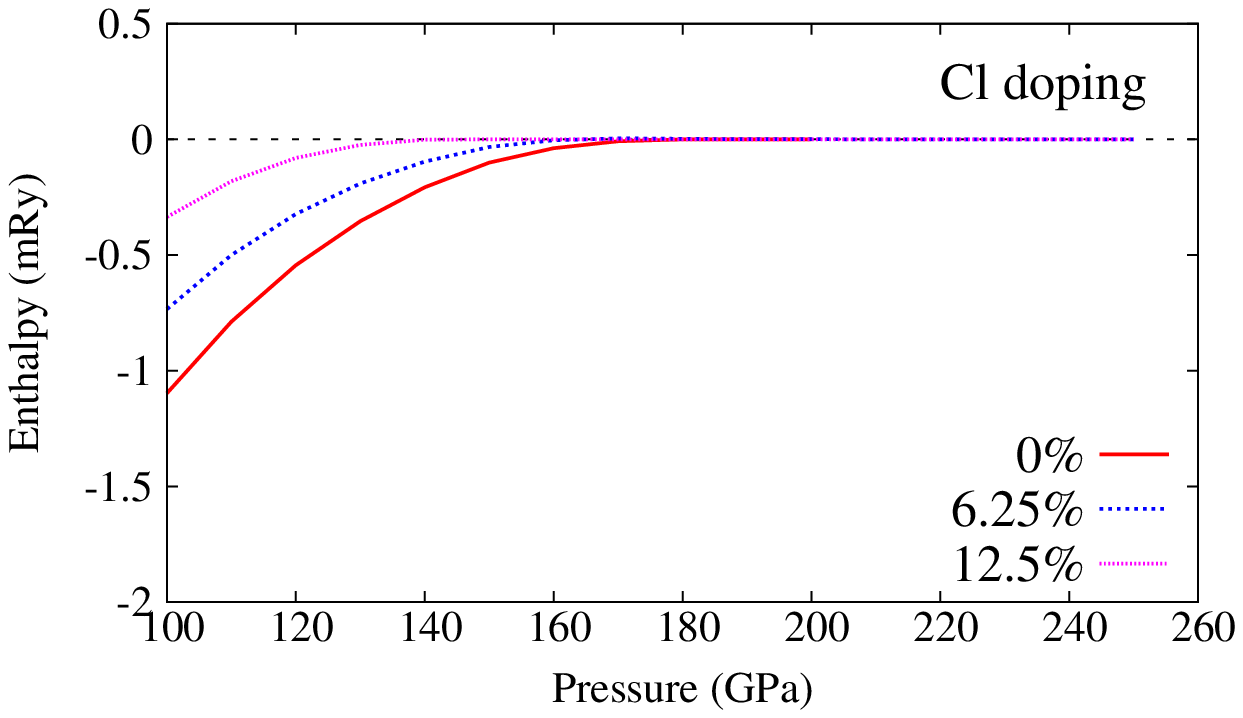} 
  \end{tabular}
  \caption{
Enthalpies of the $R3m$ phase relative to those 
of the $Im\bar{3}m$ phase for the P-doped and Cl-doped H$_3$S.
   }
\label{fig:enthalpy}
\end{center}
\end{figure}
% \end{widetext}

The x-ray diffraction measurements
 and the first-principles calculations
 including the anharmonic effect
 show that H$_3$S has a crystal structure
 with a space group of a hexagonal $R3m$
 in the pressure region above 100\,GPa
 and transforms into a cubic $Im\bar{3}m$ structure
 at around 150\,GPa,\cite{Einaga2016, Errea2016}
 in which hydrogen atoms are moved from asymmetric
 to symmetric positions between the S atoms.\cite{Duan2014}
First
 we investigated this structural phase transition in the doped system. 
Figure \ref{fig:enthalpy}
 shows the enthalpies of $R3m$
 relative to that of $Im\bar{3}m$ for H$_3$S, 
 H$_3$S$_{0.9375}$P$_{0.0625}$, H$_3$S$_{0.875}$P$_{0.125}$, 
 H$_3$S$_{0.9375}$Cl$_{0.0625}$, and H$_3$S$_{0.875}$Cl$_{0.125}$. 
Non-doped H$_3$S continuously transforms from $R3m$ into $Im\bar{3}m$ at 190\,GPa,
 which is higher by about 40\,GPa
 than the transition pressure observed experimentally
but is almost consistent with that obtained using first-principles calculations
 without the anharmonic effect.\cite{Duan2014}
The transition pressure shows no change
 with P doping,
 whereas it shifts to lower pressure
 with the increase in Cl doping,
 \textit{i.e.},
 from 190\,GPa for H$_3$S
 to 160\,GPa for H$_3$S$_{0.9375}$Cl$_{0.0625}$
 and to 140\,GPa for H$_3$S$_{0.875}$Cl$_{0.125}$.

\begin{figure}[htbp]
\begin{center}
  \includegraphics[width=8.5cm]{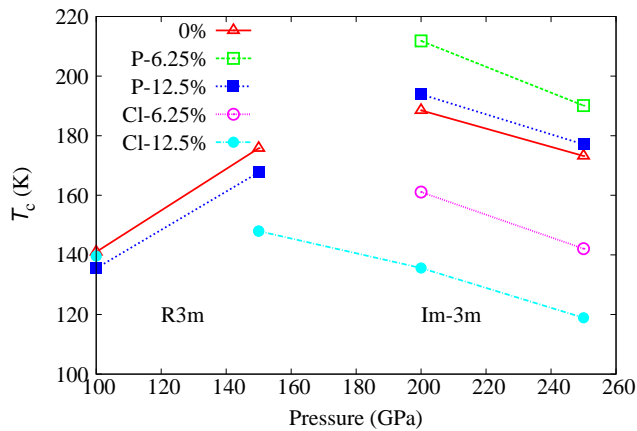}
  \caption{
    Pressure dependence of $T_\mathrm{c}$ for the doped H$_3$S. 
    The transition pressures from $R3m$ into $Im\bar{3}m$ are as follows: 190\,GPa 
    for H$_3$S, H$_3$S$_{0.9375}$P$_{0.0625}$, and H$_3$S$_{0.875}$P$_{0.125}$, 160\,GPa 
    for H$_3$S$_{0.9375}$Cl$_{0.0625}$, and 140\,GPa for H$_3$S$_{0.875}$Cl$_{0.125}$.
}
  \label{fig:tcp}
\end{center}
\end{figure}

Figure \ref{fig:tcp}
 shows the pressure dependence of the $T_\mathrm{c}$
 calculated by the Allen-Dynes formula
 for H$_3$S, H$_3$S$_{0.9375}$P$_{0.0625}$, H$_3$S$_{0.875}$P$_{0.125}$,
 H$_3$S$_{0.9375}$Cl$_{0.0625}$, and H$_3$S$_{0.875}$Cl$_{0.125}$. 
Unfortunately,
 no data of $T_\mathrm{c}$ for 6.25\% doping was obtained
 in the $R3m$ phase owing to unexpected errors
 in the phonon calculations.
The unexpected errors occur in subroutine which computes the matrices 
representing the small group of $q$ on the pattern basis. 
Therefore, we consider that the errors are caused by not the phonon instability near the transition 
from $R3m$ into $Im\bar{3}m$ but the symmetry of the structure. 
For the non-doped H$_3$S,
 $T_\mathrm{c}$ increases with pressurization in the $R3m$ phase,
 reaches the maximum at approximately 200\,GPa
 where $R3m$ transforms into $Im\bar{3}m$,
 and decreases with a further increase in pressure.
% add
The nonmonotonic behavior of $T_\mathrm{c}$ can be
 explained by the simple harmonic oscillator model.\cite{Kaplan2018}
Similar behavior is observed in 12.5\% P-doped
 and Cl-doped samples.
At 200\,GPa, in the $Im\bar{3}m$ phase,
 we found that $T_\mathrm{c}$
 increases from 189 to 212\,K at 6.25\% P doping
 and decreases to 194\,K at 12.5\% doping (Table \ref{tab:tcp}).
In the case of Cl doping,
 $T_\mathrm{c}$ decreases to 161\,K at 6.25\% doping
 and to 136\,K at 12.5\% doping. 
These results suggest that
 at low levels of P doping, \textit{i.e.}, low levels of hole doping,
 enhances the $T_\mathrm{c}$ of H$_3$S.
On the other hand, Cl doping, \textit{i.e.}, electron doping,
 decreases the $T_\mathrm{c}$
 but causes the $Im\bar{3}m$ phase to be stabilized in a lower pressure region
 compared to non-doped H$_3$S.

\begin{table}[htbp]
\caption{
    Doping dependence of 
    (a) $N(E_\mathrm{F})$,
    (b) $\lambda$, 
    (c) $\omega_{\log}$, 
    (d) $T_\mathrm{c}$ calculated using the Allen-Dynes formula, and 
    (e) $T_\mathrm{c}$ calculated directly solving
    the isotropic Eliashberg equation \cite{McMillan1968} for $Im\bar{3}m$ H$_3$S. 
}
\begin{tabular}{ccccccc}
\hline
  P(GPa) &Cl-12.5\%&Cl-6.25\%&non-doped& P-6.25\%& P-12.5\% \\ \hline  

\multicolumn{6}{l}{(a) $N(E_\mathrm{F})$ (states/eV/atom)} \\
     100 &    0.112 &        &   0.120 &         &    0.123 \\
     150 &    0.106 &        &   0.121 &         &    0.129 \\
     200 &    0.105 &  0.114 &   0.128 &   0.137 &    0.132 \\
     250 &    0.105 &  0.115 &   0.131 &   0.137 &    0.130 \\ \hline

\multicolumn{6}{l}{(b) $\lambda$} \\
     100 &     1.85&         &     1.70&         &     1.89 \\
     150 &     2.18&         &     2.22&         &     2.22 \\
     200 &     1.32&     1.52&     1.94&     2.32&     2.22 \\
     250 &     1.14&     1.23&     1.58&     1.60&     1.53 \\ \hline

\multicolumn{6}{l}{(c) $\omega_{\log}$ (K)}                  \\
     100 &      947&         &     1050&         &      895 \\
     150 &      851&         &     1000&         &      950 \\
     200 &     1380&     1380&     1230&     1180&     1110 \\
     250 &     1510&     1610&     1400&     1530&     1500 \\ \hline

\multicolumn{6}{l}{(d) $T_\mathrm{c}$ (K) [Allen-Dynes formula]}   \\
     100 &      140&         &      141&         &      136 \\
     150 &      148&         &      176&         &      168 \\
     200 &      136&      161&      189&      212&      194 \\
     250 &      119&      142&      173&      190&      177 \\ \hline

\multicolumn{6}{l}{(e) $T_\mathrm{c}$ (K) [Isotropic Eliashberg equation]} \\
     100 &      187&         &      190&         &      174 \\
     150 &      198&         &      215&         &      225 \\
     200 &      179&      199&      225&      268&      249 \\
     250 &      161&      182&      248&      251&      229 \\
\hline
\end{tabular}
\label{tab:tcp}
\end{table}
% reference
% ~/H24S7P_R3m/7elph/tc_p_dope.out

\begin{figure}[htbp]
\begin{center}
  \begin{tabular}{c}
  \includegraphics[width=8.5cm]{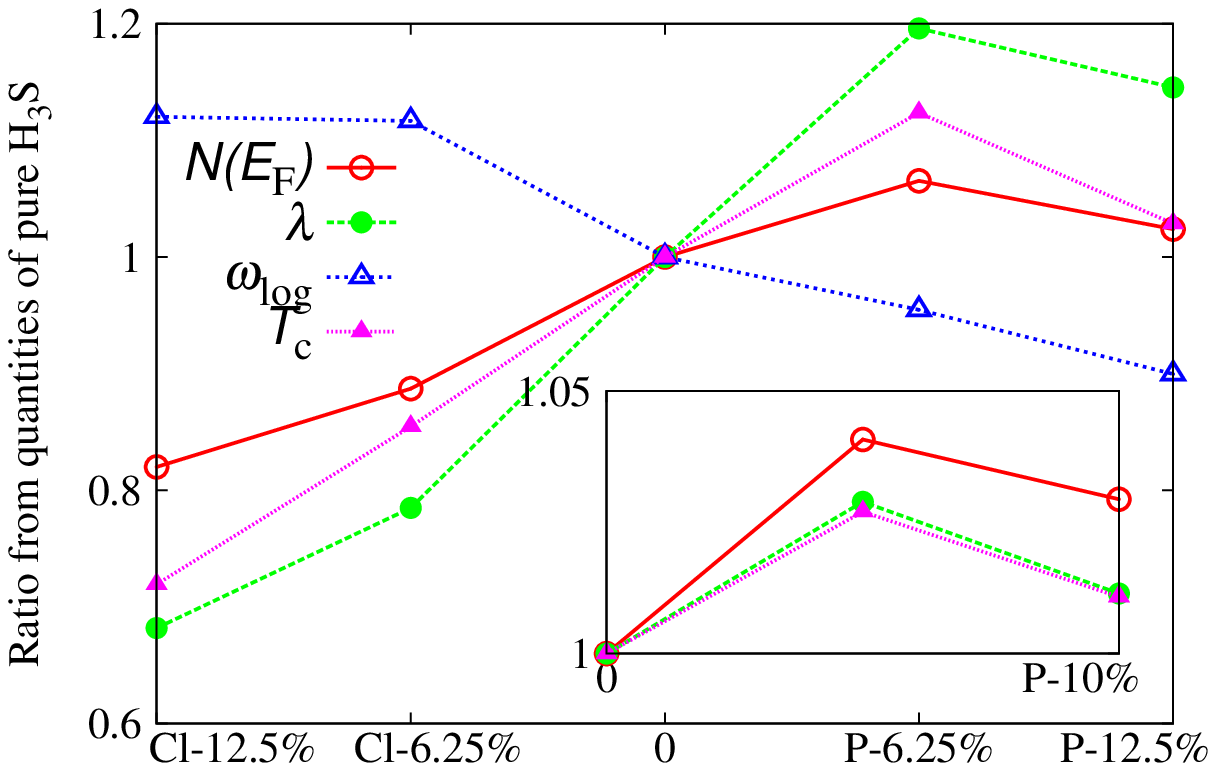} \\
  \includegraphics[width=8.5cm]{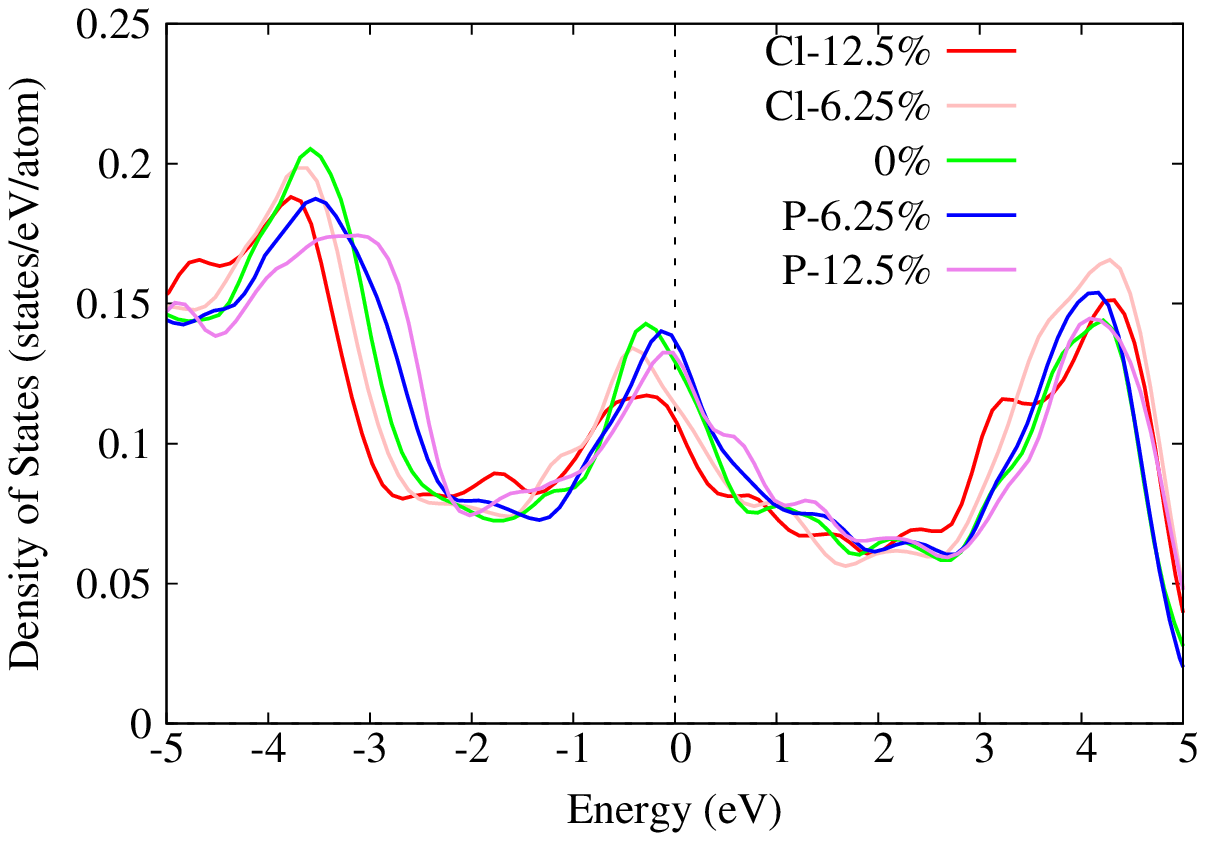} 
  \end{tabular}
  \caption{
    Doping dependence of $N(E_\mathrm{F})$, $\lambda$, $\omega_{\log}$, 
    $T_\mathrm{c}$ calculated by the Allen-Dynes formula, and DOS for $Im\bar{3}m$ 
    H$_3$S at 200\,GPa. The inset of the upper panel shows the VCA results
    reported by Fan \textit{et al.} for P doping.\cite{Fan2016}
   }
\label{fig:dlwt}
\end{center}
\end{figure}

Figure \ref{fig:dlwt}
 shows the doping dependence of $T_\mathrm{c}$, $\omega_{\log}$, $\lambda$,
 and the density of states at Fermi level $N(E_\mathrm{F})$
 in the $Im\bar{3}m$ phase at 200\,GPa
 normalized by the values of those of non-doped H$_3$S.
The original values are listed in Table \ref{tab:tcp}. 
The doping dependence of $T_\mathrm{c}$
 is explained by that of $N(E_\mathrm{F})$. 
As reported earlier,\cite{Duan2014, Fan2016}
 $Im\bar{3}m$ H$_3$S shows a large peak in DOS around the Fermi level, \textit{i.e}, 0\,eV,
 which has been suggested as a reason for the high-$T_\mathrm{c}$ observed in H$_3$S.
However,
 strictly speaking,
 the peak maximum is located at $-0.2$\,eV. 
Therefore, 
 the Fermi level is shifted toward the peak maximum of DOS with hole doping, 
 which causes a further increase of $N(E_\mathrm{F})$
 with a low P doping of 6.25\%. 
Fan \textit{et al.} investigated DOS for non-doped H$_3$S, 15\% P-doped H$_3$S, 
 and 50\% P-doped H$_3$S using the VCA calculations and found that these DOS plots have all very similar structure
 except for the position of the Fermi level (see Fig. 3 in Ref. \cite{Fan2016}). 
They suggest that the simple level shift like a rigid band model is applicable to P-doped H$_3$S and 
 the $N(E_\mathrm{F})$ maximum is obtained by the 15\% doping, 
 at which the Fermi level reaches the peak maximum of DOS. 
However,
 our supercell calculations show that 
 the doping causes not only the Fermi level shift 
 but also the peak broadening, and 
 $N(E_\mathrm{F})$ decreases with further doping 
 before the Fermi level reaches the peak maximum as observed at 12.5\% P doping
 (see the lower panel of Fig. \ref{fig:dlwt}). 
Therefore,
 we suggest that 
 the doping amount required for the $N(E_\mathrm{F})$ maximum is smaller than 
 that predicted from the simple level shift and the Fan's VCA calculations. 
In the case of electron doping,
 the Fermi level is shifted to higher energy compared to non-doped H$_3$S,
 which causes only a decrease of $N(E_\mathrm{F})$. 

% reference
% a
% ~/H24S7P_R3m/7elph/tlwd_dope.out
% ~/H24S7P_R3m/7elph/taba1.out
% b
% ~/H24S7P_R3m/5dos/deg0.020/Cl125/pwscf.pdos_tot
% ~/H24S7P_R3m/5dos/deg0.020/Cl625/pwscf.pdos_tot
%  ~/H24S7P_R3m/5dos/deg0.020/P000/pwscf.pdos_tot
%  ~/H24S7P_R3m/5dos/deg0.020/P625/pwscf.pdos_tot
%  ~/H24S7P_R3m/5dos/deg0.020/P125/pwscf.pdos_tot

Here we compare our results obtained using the supercell method
with those of the VCA method
 reported by Ge \textit{et al.} \cite{Ge2016} and by Fan \textit{et al.}.\cite{Fan2016} 
In Ge's results, at 200\,GPa,
 $T_\mathrm{c}$ is increased from 197 to 240\,K
 with 7.5\% P doping, \textit{i.e.}, a rate of $+$5.73\,K/\%,
 and is decreased from 197 to 94\,K
 with 12.5\% Cl doping, \textit{i.e.}, -8.24\,K/\%. 
In Fan's results,
 $T_\mathrm{c}$
 is increased from 185 to 190\,K with 5\% P doping,
 \textit{i.e.}, at a rate of $+$1\,K/\%. 
In contrast, in our supercell results,
 the rates are $+$3.68\,K/\% for P doping and -4.24\,K/\% for Cl doping,
 which suggests that our results are qualitatively consistent
 with the VCA results 
 but the doping effect is less (more) remarkable than that predicted by 
 Ge \textit{et al.} (Fan \textit{et al.}). 

\begin{figure}[htbp]
\begin{center}
  \includegraphics[width=8.5cm]{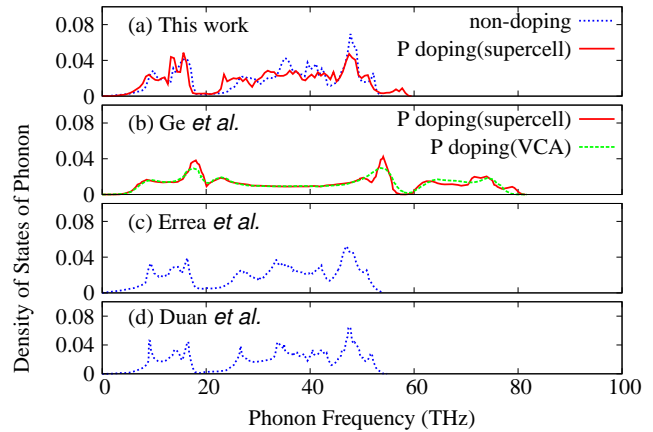}
  \caption{
    Phonon DOS for $Im\bar{3}m$ H$_3$S at 200\,GPa: 
    (a) our supercell results for non-doped and at 12.5\% P doping, 
    (b) Ge's supercell and VCA results at 12.5\% P doping,\cite{Ge2016}
    (c) Errea's results for non-doped,\cite{Errea2015} and 
    (d) Duan's results for non-doped.\cite{Duan2014}
}
  \label{fig:phdos}
\end{center}
\end{figure}

We discuss the reasons for the difference
 in the doping effects on the superconductivity
 between the VCA and our supercell results. 
The inset of Fig. \ref{fig:dlwt}
 shows the normalized $T_\mathrm{c}$, $\lambda$, and $N(E_\mathrm{F})$ values
 reported by Fan \textit{et al.}. 
The $\lambda$ value in our results
 is comparatively higher
 than that in Fan's results for P doping. 
This difference is considered to be caused
 by the difference of the phonon calculation method. 
They calculated the superconducting parameters for the doped system
 using the phonon frequency obtained from the calculated results for non-doped H$_3$S,\cite{Duan2014}
On the other hand,
 we directly calculated the phonon frequency
 using the supercell method. 
Consequently, larger $\lambda$ and higher $T_\mathrm{c}$ were obtained in our calculations
 in comparison to  Fan's calculations. 
Figure \ref{fig:phdos}
 shows a comparison of the phonon DOS of $Im\bar{3}m$ H$_3$S at 200\,GPa
 among the present work and previous reports.\cite{Duan2014, Errea2015}
Comparing our result at 12.5\% P doping [Fig. \ref{fig:phdos}(a)]
 with that of Ge's [Fig. \ref{fig:phdos}(b)],
 we found that the maximum of the phonon frequency in our phonon DOS is lower by about 20\,THz.
To judge the calculation accuracy,
 we also compared our phonon DOS with
 those of non-doped H$_3$S previously reported
 by Duan \textit{et al.} \cite{Duan2014} and Errea \textit{et al.}.\cite{Errea2015}
Since the phonon DOS is shown as a projection on each atom in their papers,
 we extracted the data from those reports,
 obtained the total DOS by summing up the data,
 and plotted in Figs. \ref{fig:phdos} (c) and (d). 
Consequently,
 we found that our phonon DOS for the non-doped H$_3$S shows
 a good agreement with those reported earlier. 
The maximum of the phonon frequency is shifted by 5.7\,THz
 towards higher frequency with P doping,
 whereas the phonon DOS is not expected to be drastically changed. 
Therefore,
 we conclude that our results on the phonon DOS and doping effect
 are more reliable than those calculated by Ge \textit{et al.}.

Finally we discuss on the feasibility of P-doped H$_3$S. 
We calculated the defect formation enthalpy $H_\mathrm{f}$ 
 for H$_3$S$_{0.9375}$P$_{0.0625}$ and H$_3$S$_{0.875}$P$_{0.125}$
 according to the following equation:
\begin{eqnarray}
  H_\mathrm{f}
  &=& 
  H_\mathrm{H_3S_{1-\it{x}}P_{\it{x}}} + xH_\mathrm{S}
 -H_\mathrm{H_3S} - xH_\mathrm{P}
\end{eqnarray}
 where $H_\mathrm{H_3S_{1-\it{x}}P_{\it{x}}}$ and $H_\mathrm{H_3S}$
 represent the enthalpies of H$_3$S$_{1-x}$P$_{x}$ ($x$ = 0.0625, 0.125)
 and non-doped H$_3$S, respectively.
$H_\mathrm{S}$ and $H_\mathrm{P}$ are
 the enthalpies per atom for S and P elements. 
Table \ref{tab:tcp1} shows $H_\mathrm{f}$ per atom in temperature units. 
All the doped systems are thermodynamically unstable compared with non-doped H$_3$S, 
 whereas the enthalpy differences show few hundreds kelvin,
 \textit{e.g.} 167.3\,K at 100\,GPa for 6.25\% P-doping. 
Very recently,
 Guigue \textit{et al.} and Goncharov \textit{et al.}
 directly synthesized pure H$_3$S from S and H$_2$
 in laser-heated ($<$ 1300\,K) diamond anvil cells
 under high pressure.\cite{Guigue2017,Goncharov2017}
Therefore,
 we suggest that P-doped H$_3$S can be synthesized as metastable states 
 by similar laser-heating experiments
 in the mixture of S, H$_2$, and a small amount of P. 

\begin{table}[htbp]
\caption{
    Formation enthalpy of H$_3$S$_{1-x}$P$_{x}$.
    Abbreviations, rh, sc, and sh, represent 
    a rhombohedral structure of $\beta$-Po type,
    a simple cubic structure, and
    a simple hexagonal structure, respectively. 
 }
\begin{tabular}{cccccc}
\hline
    $P$ & \multicolumn{2}{c}{structure} & $x$ & \multicolumn{2}{c}{$H_\mathrm{f}$}  \\  
  (GPa) & S  & P   &        & (mRy/atom)  & (K) \\ \hline
  100   & rh & sc  & 0.0625 & 1.060       & 167.3 \\
        &    &     & 0.125  & 1.759       & 277.7 \\
  150   & rh & sh  & 0.0625 & 1.367       & 215.9 \\ 
        &    &     & 0.125  & 2.297       & 362.7 \\
  200   & rh & sh  & 0.0625 & 1.508       & 238.0 \\ 
        &    &     & 0.125  & 2.683       & 423.6 \\
  250   & rh & sh  & 0.0625 & 1.523       & 240.5 \\ 
        &    &     & 0.125  & 2.831       & 447.0 \\
\hline
\end{tabular}
\label{tab:tcp1}
\end{table}
% reference
% ~/P/3enthalpy/job/formation.out_6.25P
% ~/P/3enthalpy/job/formation.out_12.5P
% ~/P/sc/3enthalpy/job/formation.out_6.25P
% ~/P/sc/3enthalpy/job/formation.out_12.5P

\section{Summary}
In this study,
 we investigated the doping effect
 on the superconductivity of $Im\bar{3}m$ H$_3$S
 using first-principles calculations
 based on the supercell method, which gives more reliable results on the superconductivity in doped systems
 than the calculations
 based on the virtual crystal approximation reported earlier. 
First
 we explored the pressure-induced structural phase transition from 
$R3m$ to $Im\bar{3}m$ phase in the doped system and 
found that the transition pressure shows no change with P doping, 
whereas it is shifted towards lower pressure by Cl doping, \textit{i.e.}, from 190 to 160 (140)\,GPa 
with 6.25 (12.5)\% doping. 
These results suggest that the $Im\bar{3}m$ H$_3$S phase with high-$T_\mathrm{c}$,
 which is experimentally observed in the pressure region above 150\,GPa,
 can be observed at lower pressure with electron doping. 
In the $Im\bar{3}m$ phase, at 200\,GPa,
 $T_\mathrm{c}$ is increased
 from 189 to 212\,K with 6.25\% P doping
 at a rate of $+$3.68\,K/\%,
 and then decreased to 192\,K by over doping to 12.5\%. 
In contrast,
 $T_\mathrm{c}$ decreases with the increase of Cl doping;
 161\,K at 6.25\% and 136\,K at 12.5\%. 
The doping dependence of $T_\mathrm{c}$
 is mainly explained by the $N(E_\mathrm{F})$ with doping. 
Low levels of P doping, \textit{i.e.}, low levels of hole doping,
 causes the Fermi level towards the peak maximum of DOS increasing $N(E_\mathrm{F})$. 
However,
 over doping, such as 12.5\% doping causes the peak itself to weaken, decreasing $N(E_\mathrm{F})$. 
In the case of Cl doping, \textit{i.e.}, the electron doping,
 the Fermi level is shifted towards higher energy
 compared to non-doped H$_3$S, which causes a decrease in $N(E_\mathrm{F})$. 
These trends are qualitatively consistent with
 the VCA results reported earlier,
 whereas the $T_\mathrm{c}$ values are different from each other 
 owing to the difference of the phonon calculation methods; 
 a rate of $+$5.73\,K/\% and $+$1.00\,K/\%
 according to Ge's and Fan's results, respectively. 
In conclusion,
 our supercell calculations suggest
 that the superconductivity of $Im\bar{3}m$ H$_3$S
 can be enhanced further by low levels of hole doping 
 showing an increase of about 20\,K with 6.25\% P doping. 
The formation enthalpy of 6.25\% P-doped H$_3$S is about few hundreds kelvin, 
which can be synthesized from S, H$_{2}$, and a small amount of P by 
laser-heating experiments.

\begin{acknowledgments}
This work was supported by JSPS KAKENHI, under Grant-in-Aid for Specially Promoted Research (26000006) 
and Grant-in-Aid for Scientific Research (C) (17K05541), 
 Asahi Glass Foundation, Yamada Science Foundation, 
and MEXT under ''Exploratory Challenge on Post-K computer'' (Frontiers of Basic Science: Challenging the Limits).
\end{acknowledgments}

\end{document}